\documentclass[aps, twocolumn, prl, floatfix, superscriptaddress,longbibliography]{revtex4-2}
\usepackage{amsmath,amssymb,amsthm}
\usepackage{graphicx}
\usepackage{xcolor}

\usepackage{color}
\definecolor{darkblue}{rgb}{0,0,.65}
\definecolor{darkgreen}{rgb}{0.3,0.6,0.3}
\definecolor{cyan1}{rgb}{0.0, 0.6, 0.6}
\usepackage[%
pdfstartview=FitH,%
breaklinks=true,%
bookmarks=true,%
colorlinks=true,%
anchorcolor=black,%
citecolor=red,
filecolor=black,%
menucolor=black,%
urlcolor=darkblue,%
linkcolor=blue,%
]{hyperref}
\usepackage{cleveref}  
\usepackage{orcidlink}

\usepackage{tikz}
\usetikzlibrary{quantikz2}

\begin{document}
 
\title{Manipulation of Topological Corner States via Subchiral Symmetry}
\author{Hai-Tao Ding}
\thanks{These authors contributed equally to this work.}
\email{htding.9@nus.edu.sg}
\affiliation{Centre for Quantum Technologies, National University of Singapore, 3 Science Drive 2, Singapore 117543}
\affiliation{MajuLab, CNRS-UNS-NUS-NTU International Joint Research Unit, Singapore UMI 3654, Singapore}

\author{Tianqi Chen\,\orcidlink{0000-0002-2336-1875 }}
\thanks{These authors contributed equally to this work.}
\email{chen\_tianqi@a-star.edu.sg}
 \affiliation{Bioinformatics Institute, Agency for Science, Technology and Research (A*STAR), 30 Biopolis Street, No.~07-01 Matrix, Singapore 138671\looseness=-1}
 \affiliation{Institute of High Performance Computing (IHPC), Agency for Science, Technology and Research (A*STAR),
1 Fusionopolis Way, No.~16-16 Connexis, Singapore 138632\looseness=-1}

\author{Leong-Chuan Kwek}
\email{kwekleongchuan@nus.edu.sg}
\affiliation{Centre for Quantum Technologies, National University of Singapore, 3 Science Drive 2, Singapore 117543}
\affiliation{MajuLab, CNRS-UNS-NUS-NTU International Joint Research Unit, Singapore UMI 3654, Singapore}
\affiliation{National Institute of Education, Nanyang Technological University, Singapore 637616, Singapore}
\affiliation{Quantum Science and Engineering Centre (QSec), Nanyang Technological University, Singapore}

\author{Jiangbin Gong}
\email{phygj@nus.edu.sg}
\affiliation{Centre for Quantum Technologies, National University of Singapore, 3 Science Drive 2, Singapore 117543}
\affiliation{Department of Physics, National University of Singapore, 2 Science Drive 3, Singapore 117551, Singapore}
\affiliation{MajuLab, CNRS-UNS-NUS-NTU International Joint Research Unit, Singapore UMI 3654, Singapore}

\begin{abstract}
Higher-order topological phases provide robust corner modes, but their use requires controllable creation, isolation, and transfer of individual modes and their superpositions. Here we demonstrate, using the two-dimensional Benalcazar–Bernevig–Hughes model as an example, that subchiral symmetry provides a general control principle for manipulating topological corner modes. The conventional chiral symmetry decomposes into four subchiral symmetries, each associated with one zero-energy corner mode. By selectively breaking these subsymmetries with controlled intercell hoppings, we reduce the fourfold corner-state manifold step by step to single isolated modes. We further design adiabatic protocols that transfer either a single corner state or a superposition of two corner states between selected corners, while preserving the relative phase in the latter case. Both numerical simulations and IBM quantum-processor implementations show that the proposed protocols can be executed with high fidelity, establishing subchiral symmetry as a route to programmable higher-order topological state manipulation.
\end{abstract}

\maketitle


\textit{Introduction.}--
 Symmetry-protected boundary modes hosted by topological lattices  are robust against local perturbations~\cite{hasan2010colloquium,qi2011topological,bernevig2013topological,zhang2018topological,armitage2018weyl,cooper2019topological,hasan2011three,shen2011topological,tokura2019magnetic,moore2010birth} and hence appealing for information encoding, quantum state transfer and quantum computation~\cite{kitaev2003fault,nayak2008non,alicea2012new,PhysRevLett.113.087403,beenakker2013search,zhou2014high,sarma2015majorana,mittal2016topologically,mei2018robust,lutchyn2018majorana,yan2018majorana,PhysRevLett.125.150502,pahomi2020braiding,bomantara2020measurement,minganti2021dissipative,wang2022arbitrary,tan2020high,arkhipov2024restoring}. In higher-order topological insulators, topological boundary modes are localized at lower-dimensional boundaries, such as hinges and corners~\cite{benalcazar2017quantized,langbehn2017reflection,benalcazar2017electric,song2017d,Schindler2018SA,ezawa2018higher,geier2018second,khalaf2018higher,van2018higher,trifunovic2019higher,jia2023unified}, as realized in a variety of experimental platforms~\cite{peterson2018quantized,serra2018observation,imhof2018topolectrical,xue2019acoustic,mittal2019photonic,Dutt2020LSA,noguchi2021evidence,wu2023higher,xu2023observation,deng2024high}.   However, the existence of topological corner modes or hinge modes alone is not sufficient to facilitate topology-based quantum control: one must be able to selectively create, isolate, and transport such states, including their superpositions. 

Despite rapid progress in quantum-state manipulation on superconducting qubits, trapped ions, and photonic platforms~\cite{yung2005perfect,burgarth2005conclusive,sillanpaa2007coherent,Li2018Perfect,ge2025enhancedalgorithmicperfectstate,rowe2002transportquantumstatesseparation,Chen2024geometry,stute2013quantum,chapman2016experimental,kurpiers2018deterministic,Maddox2024enhanced,zhang2025topological,wang2026observation}, 
a symmetry-based control principle for selectively manipulating and phase-coherently transferring topological corner states is still lacking.
Notably, thanks to the concept of  subchiral symmetry introduced as a refined symmetry principle for resolving zero modes in chiral-symmetric topological systems~\cite{wang2023sub,verma2024non,kang2024subsymmetry}, to address individual boundary modes within a degenerate zero-mode manifold without destroying their topological protection has just become plausible. 
 Specifically, in the Su--Schrieffer--Heeger chain and related settings,  subchiral symmetry can decompose conventional chiral protection into symmetry sectors associated with individual edge states, allowing selected zero modes to be removed by breaking the corresponding subchiral symmetry~\cite{su1979solitons,wang2023sub}. 
 Going beyond using subchiral symmetry as a diagnostic tool for zero-mode counting and selective elimination,  leveraging subchiral symmetry for robust and phase-coherent transport of subchiral-symmetry-protected corner states is not only of theoretical interest to dynamical studies of higher-order topological matter, but also makes the corresponding high-fidelity implementation on quantum platforms possible under a realistic setting.

Below we use subchiral symmetry as a dynamical control principle for topological corner state transfer. We generalize subchiral symmetry~\cite{wang2023sub,verma2024non,kang2024subsymmetry} to the two-dimensional (2D) Benalcazar-Bernevig-Hughes (BBH) model, which in the topological regime  is a paradigmatic second-order topological insulator (SOTI) hosting four zero-dimensional corner modes under open boundary conditions~\cite{benalcazar2017quantized}. The chiral symmetry of the BBH Hamiltonian can be decomposed into four subchiral symmetries, each associated with one of these corner modes. By selectively breaking these symmetries with engineered intercell hoppings, the fourfold zero-mode manifold can be reduced step by step to isolated corner modes. This symmetry-resolved control enables adiabatic transfer of both a single corner state and a two-corner-state superposition between selected corners.  In the superposition protocol, even the relative phase can be preserved, confirming coherent transfer within the degenerate zero-mode manifold. To understand how our topological corner states and their superpositions respond to our control protocols under a noisy environment, we execute the protocols on an IBM superconducting quantum processor and showcase their robustness, with all the examples tested yielding high-fidelity outcomes.


\textit{SOTI with Sub-symmetry.}--
\begin{figure}[t]
	\centering
\includegraphics[width=1.0\columnwidth,draft=false]{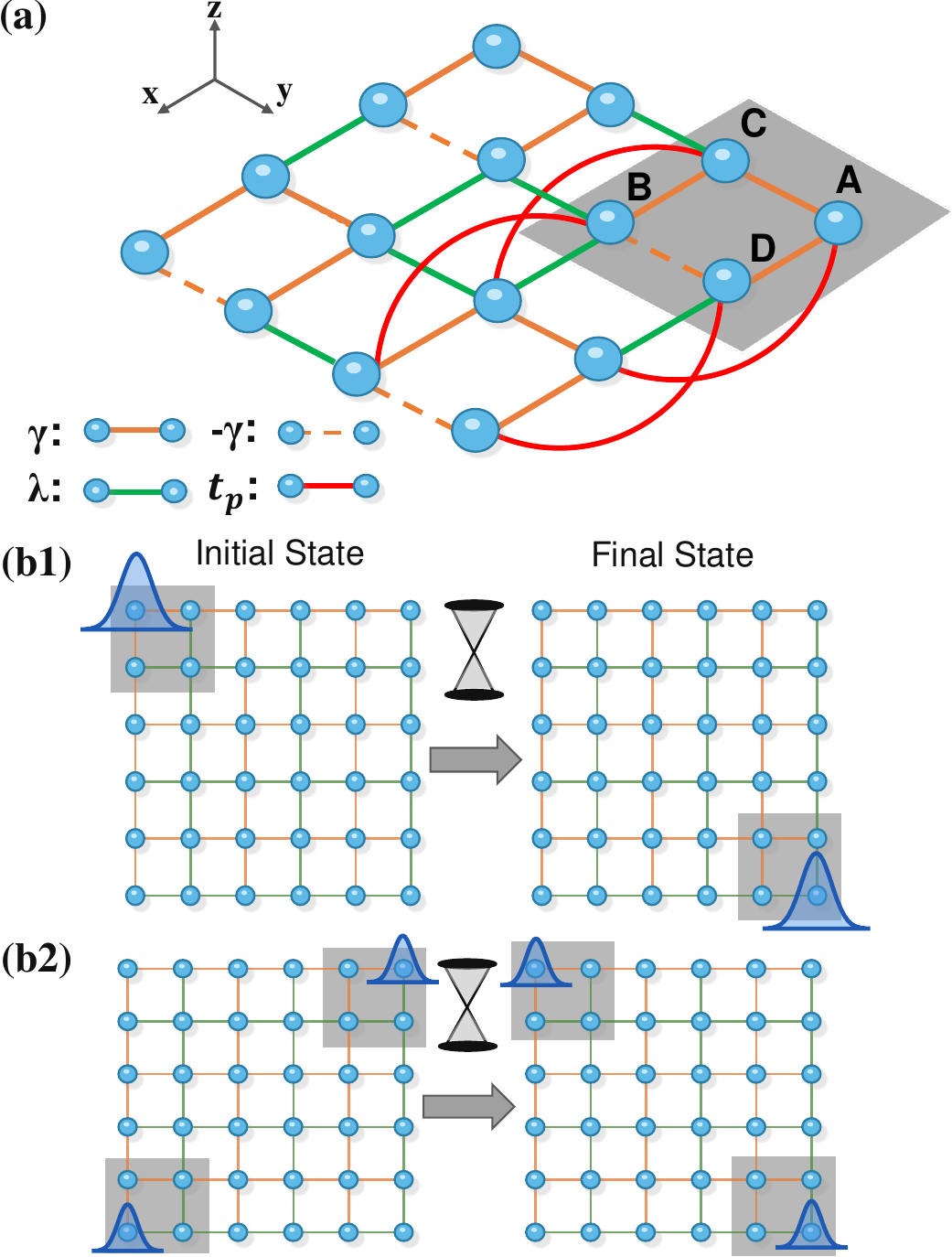}
	\caption{Subchiral-symmetry-controlled manipulation of corner modes in an extended 2D BBH model. (a) Lattice geometry and hopping configuration. The unit cell contains four orbitals, labeled A, B, C, and D. The conventional BBH hoppings are denoted by $\pm\gamma$ and $\lambda$, while the red links represent controllable hoppings between identical orbitals in adjacent unit cells along the $x$ direction, denoted by $t_p$. These hoppings selectively break the corresponding subchiral symmetries. (b1) Adiabatic transfer of a single isolated corner mode. (b2) Adiabatic transfer of an equal-weight superposition of two corner modes between opposite-corner pairs. Blue envelopes denote the corner-state probability distributions.}
	\label{fig: BBH illustration}
\end{figure} 
We consider a 2D BBH model~\cite{benalcazar2017quantized} as depicted in Fig~\ref{fig: BBH illustration}. The tight-binding Hamiltonian is 
\begin{equation}
H = \sum_{\mathbf{R}} \Psi_{\mathbf{R}}^\dagger H_{0} \Psi_{\mathbf{R}} + \sum_{\mathbf{R},\mu=x,y} \left( \Psi_{\mathbf{R}+\mu}^\dagger T_{\mu} \Psi_{\mathbf{R}} + \text{H.c.} \right), 
\label{BBH_model}
\end{equation}
where $\Psi_{\mathbf{R}} = (c_{\mathbf{R},A}, c_{\mathbf{R},B}, c_{\mathbf{R},C}, c_{\mathbf{R},D})^T$ is a four-component annihilation operator 
at the unit cell $\mathbf{R} = (x,y)$. The matrices $H_0$, $T_x$, and $T_y$ represent the intra-cell and inter-cell hopping processes along $x$ and $y$ directions, respectively, with $H_0 = \gamma \Gamma_4 + \gamma \Gamma_2$, $T_x = \frac{\lambda}{2} (\Gamma_4 + i\Gamma_3)$, and $\quad T_y = \frac{\lambda}{2} (\Gamma_2 + i\Gamma_1)$.
Here $\Gamma_i=-\tau_2\sigma_i$ ($i=\{1,2,3\}$), and $\Gamma_4=\tau_1\sigma_0$, with $\tau,\sigma$ are Pauli matrices. 
This Hamiltonian preserves the time-reversal symmetry, charge-conjugation symmetry, and chiral symmetry. The chiral symmetry operator is $\Gamma=\tau_3\sigma_0$ 
with $\Gamma H\Gamma^{-1}=-H$.  For $|\gamma / \lambda|<1$, it is a second-order topological insulator phase and a trivial insulator when $|\gamma / \lambda|>1$, characterized by the quantized bulk quadrupole moment~\cite{benalcazar2017quantized} and multipole chiral numbers~\cite{benalcazar2022chiral} whereas the polarization and the Chern number are zero~\cite{benalcazar2017quantized,benalcazar2022chiral,SuppleMat}.
Throughout this work, we set $\lambda=1$. All energies are measured in units of $\lambda$, and time is measured in units of $1/\lambda$.

\begin{figure*}[t]
	\centering
\includegraphics[width=1.0\textwidth]{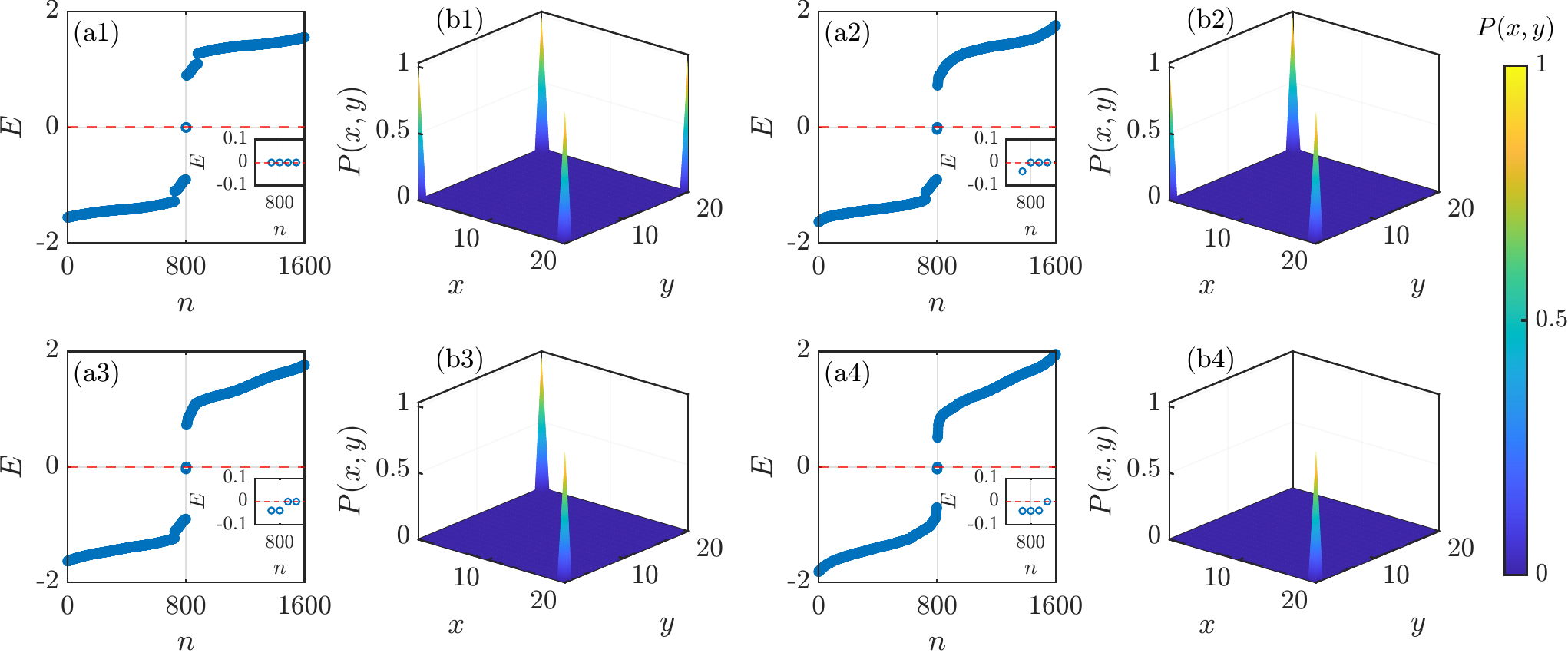}
	\caption{(a1-a4) The energy spectrum and (b1-b4) the spatial distributions $P(x,y)$ for the zero-energy corner states. Here $(x, y)$ denotes the unit-cell position, and $P(x,y)$ is the unit-cell-resolved probability density obtained by summing the wave-function weight over the four sublattices $A$, $B$, $C$, and $D$. The insets of (a1-a4) shows a zoom-in of the spectrum around $E=0$. (a1) The energy spectrum of $H$ in Eq.~\eqref{BBH_model}, showing four degenerate zero-energy corner states isolated within the bulk and edge gaps. (b1) The corresponding distribution $P(x,y)$, showing localization at the four corners. (a2)--(a4) Energy spectra for $H+H_A$, $H+H_A+H_B$, and $H+H_A+H_B+H_C$, respectively, where the zero-mode manifold is successively reduced to three, two, and one corner state. (b2)--(b4) The corresponding distributions $ P(x,y)$, showing that the remaining zero modes are localized at the corresponding corners. The calculations are performed on a lattice with $N_x \times N_y = 20 \times 20$ unit cells. Other parameters are $\lambda=1$, $\gamma=0.1$, $t_p=0.2$.}
\label{fig:BBH_corner_state}
\end{figure*}

The energy spectrum of the Hamiltonian under open boundary conditions (OBC) in both $x$ and $y$ directions is shown in Fig.~\ref{fig:BBH_corner_state}(a1). The spectrum exhibits four degenerate zero-energy corner states isolated within the bulk and edge gaps. The spatial distributions of  the four zero-energy modes, localized at the four corners of the lattice respectively, are illustrated in Fig.~\ref{fig:BBH_corner_state}(b1), reflecting the bulk-corner correspondence of the 2D SOTI. In the BBH model, however, the chiral symmetry can be further decomposed into four subchiral symmetries associated with the four sublattices.  That is, let 
$\Gamma=P_A+P_B-P_C-P_D$, where $P_{\alpha}$ ($\alpha=A,B,C,D$) denotes the projector onto sublattice $\alpha$, with $P_A=\text{diag}(1,0,0,0)$, $P_B=\text{diag}(0,1,0,0)$, $P_C=\text{diag}(0,0,1,0)$, $P_D=\text{diag}(0,0,0,1)$. Then it is straightforward to identify four subchiral symmetries defined by $\Gamma H\Gamma^{-1}P_{\alpha}=-HP_{\alpha}$. Thus, even when the full chiral symmetry is broken, a given corner state can remain protected provided that the corresponding projected relation is preserved.

To manipulate the number of zero-energy corner states, we introduce four control terms,
\begin{equation}
    H_{\alpha} = t_p \sum_{x,y} c^{\dagger}_{x+1,y,\alpha} c_{x,y,\alpha} + \text{H.c.},
\end{equation}
with $\alpha=\{A,B,C,D\}$. Each control term $H_{\alpha}$  selectively breaks the corresponding subchiral symmetry and lifts the associated corner state away from zero energy. This produces a stepwise, symmetry-resolved reduction of the zero-mode manifold, as shown in Fig.~\ref{fig:BBH_corner_state}. 
After adding $H_A$, the A-sector corner mode is lifted, leaving three zero-energy modes localized at the remaining corners [Fig. ~\ref{fig:BBH_corner_state}(a2) and ~\ref{fig:BBH_corner_state}(b2)]. Further adding $H_B$ reduces the zero-mode manifold to two corner states [Fig. ~\ref{fig:BBH_corner_state}(a3) and ~\ref{fig:BBH_corner_state}(b3)], while adding $H_A + H_B + H_C$ gives a single isolated zero-energy corner state [Fig. ~\ref{fig:BBH_corner_state}(a4) and ~\ref{fig:BBH_corner_state}(b4)]. When all four control terms are turned on, the entire zero-energy corner manifold is removed. These spectra and spatial profiles establish a one-to-one correspondence between broken subchiral symmetries and removable corner modes. For a single control term $H_\alpha$, the associated corner mode is shifted to $E_\alpha=-2\gamma\lambda t_p/(\lambda^2+t_p^2)$, while the remaining corner modes stay pinned at zero energy up to exponentially small finite-size corrections. Analytical derivations of the orbital-resolved corner-mode wave functions and their energy shifts under the control terms are provided in the Supplementary Materials~\cite{SuppleMat}. There, the results are summarized for all ($2^4=16$) possible control configurations, showing explicitly how the number of zero-energy corner modes can be tuned from four to zero by choosing the control set~\cite{SuppleMat}.
\begin{figure*}[t]
	\centering
\includegraphics[width=1.0\textwidth]{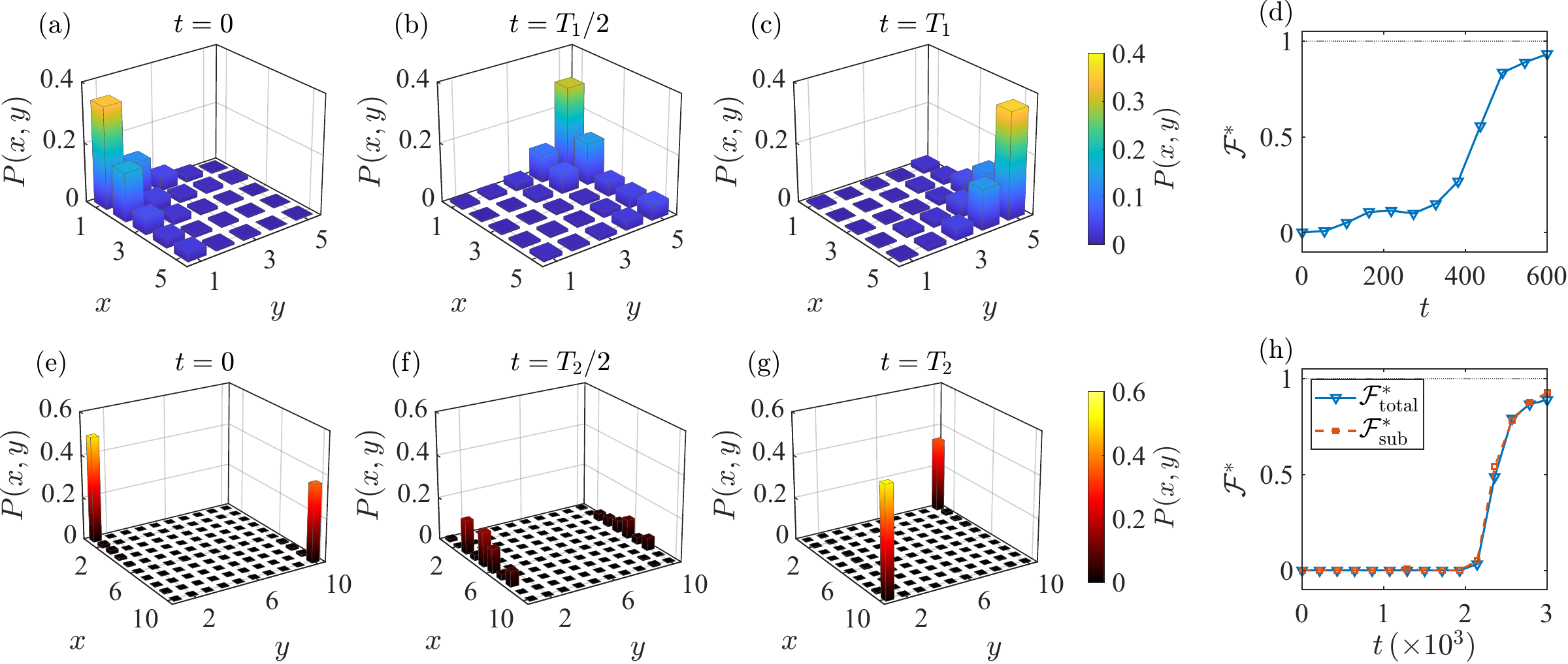}
	\caption{Transfer of corner states on the IBM quantum processor \texttt{ibm\_boston} (hardware details in the Supplementary Materials~\cite{SuppleMat}). (a)--(d) Single corner mode; (e)--(h) superposition of corner modes. (a) and (e) Measured density distributions of the initial states at $t = 0$ for the single-mode and superposition protocols, respectively. (b) and (f) Density distribution of the instantaneous state at $t=T_1/2$ and $t=T_2/2$, respectively. (c) and (g) Density distribution of the final state at $t=T_1$ and $t=T_2$, respectively. (d) Transformed fidelity $F^{*}$ for the single-mode transfer and (h) transformed total fidelity $F_{\text{total}}^{*}$ and subspace fidelity $F_{\text{sub}}^{*}$ for the superposition transfer, all approaching unity at the end of the evolution. On the hardware, the fidelity is obtained by applying an additional unitary rotation to all occupied quantum states; for the superposition case [(h)] we define and measure both subspace and total-space fidelities to characterize the manipulation. We probe the fidelity $\mathcal{F}^*$ with $12$ data points in (d) and $15$ in (h). The lattice has $N_x\times N_y=5\times5$ unit cells for (a)--(d) and $10\times10$ unit cells for (e)--(h). Other parameters: $\lambda=1$, $\gamma=0.6$, $t_p=0.2$, $T_1=600$, and $T_2=3000$.
    }
\label{fig:fig2_State_Transfer_Tianqi}
\end{figure*}

\textit{Adiabatic Transfer of Single Corner Modes.}--
We now investigate how to perform the adiabatic transfer of zero-energy corner modes via subchiral-symmetry engineering. Specifically,
we consider the following time-dependent Hamiltonian
\begin{equation}
H_{\text{tot}}(t)=H+\mathbf{g}(t)\cdot\mathbf{H_{\alpha}},
\end{equation}
where $\mathbf{g}(t)=(g_1,g_2,g_3,g_4)$ are time-dependent control functions, and $\mathbf{H_{\alpha}}=(H_A,H_B,H_C,H_D)$.
To transfer one corner mode on the topological lattice, the control functions are chosen as
\begin{equation}
\mathbf{g}(t)=
\begin{cases}
(t_p, \frac{2t_p}{T_1}t,-\frac{2t_p}{T_1}t+t_p,t_p), & t\in[0,T_1/2),\\
(-\frac{2t_p}{T_1}t+2t_p, t_p, \frac{2t_p}{T_1}t-t_p, t_p), & t\in[T_1/2, T_1].
\end{cases}
\label{eq:gcompact}
\end{equation}
The protocol is designed such that the Hamiltonian evolves continuously along a designated path in parameter space without closing the instantaneous gap. The resulting single-corner-mode transfer is schematically illustrated in Fig.~\ref{fig: BBH illustration}(b1). The initial Hamiltonian is $H_{\text{tot}}(t=0)=H+H_{A}+H_{C}+H_{D}$, hosting a single zero-energy corner state localized at the initial corner, which we take as the initial state. Under the adiabatic evolution generated by $H_\text{tot}(t)$, 
the state is first transferred to the corner state of the intermediate Hamiltonian $H_{\text{tot}}(T_1/2) = H + H_A + H_B + H_D$, and finally reaches the target corner state of $H_{\text{tot}}(T_1) = H + H_B + H_C + H_D$.

In an actual platform, there will be always noise brought in by the control protocol and from the environment. It is hence of general interest to  implement the above protocol on a programmable IBM superconducting quantum processor (ibm\_boston).  The exact Hamiltonian dynamics and noiseless circuit benchmarks are provided in the Supplementary Materials~\cite{SuppleMat}.
Consistent with the ideal adiabatic dynamics, the physical realization of the quantum state manipulation on the IBM quantum processor confirms the robustness of the sub-chiral-symmetry--enabled state manipulation. To efficiently realize such a Hamiltonian time evolution in the number-conserving sector, we employed a variational quantum algorithm to approximate the time evolution operator with a shallow circuit ansatz that is suitable for NISQ-era devices (Details of the implementation as well as the IBM hardware can be found in Supplementary Materials~\cite{SuppleMat}). In Fig.~\ref{fig:fig2_State_Transfer_Tianqi}(a-c) the hardware results are shown corresponding to the initial, the intermediate and the final stages of the density distribution of the time evolution. Despite the presence of device noise and readout errors, the measured density distributions qualitatively reproduce the expected spatial features almost the same as the noiseless results~\cite{SuppleMat}: the initial state ($t=0$) remains strongly localized at the $(1,1)$ corner of the space, the mid-evolution state ($t=T_1/2$) exhibits a localized profile at the $(1,5)$ corner, whereas the final state ($t=T_1$) is at the $(5,5)$ corner of the lattice. To quantify the transfer, we define the gauge-invariant fidelity which is the overlap between the target state ($\Psi_{\text{tar}}$) and the time-evolved state $\mathcal{F}(t)=|\langle\psi(t)|\Psi_{\text{tar}}\rangle|^2$. For a consistent and systematic comparison of the fidelity with the physical realization on the IBM hardware, instead of computing the original fidelity $\mathcal{F}\left(t\right)$ which requires quantum state tomography and is tedious even on a NISQ-era hardware~\cite{Choo2018,Chen2019}, we define the transformed fidelity $\mathcal{F}^{\ast}$ which can be easily evaluated by applying supplementary unitary rotations to all populated quantum states. This is equivalent to the original fidelity definition of $\mathcal{F}\left(t\right)$.
It is found that the final state fidelity for hardware realization shows similar tendency toward being unity. Again, for the transformed fidelity $\mathcal{F}^{\ast}$, up to the device readout error as well as the IBM native $R_{zz}$ gates infidelity, $\mathcal{F}^{\ast}$ is still larger than $91\%$. These observations demonstrate that the symmetry-guided protocol survives realistic imperfections and preserves the essential dynamical pathway underlying the corner-state transfer. Together with the measured density distribution and the fidelity with the target state, the robust hardware results have demonstrated direct evidence that sub-chiral-symmetry control offers a viable route toward experimentally realizable topological state manipulation on near-term quantum devices.

\textit{Coherent Adiabatic Transfer of Corner-Mode Superposition.}--
{Our protocol can be naturally extended to the transfer of coherent superpositions of corner modes.} 
For the control path $\mathbf{g}^{\prime}(t)=[t_p t/T_2,\, t_p t/T_2,\, -t_p t/T_2+t_p,\, -t_p t/T_2+t_p]$, the initial Hamiltonian $H_{\text{tot}}^{\prime}(t=0) = H + H_C + H_D$ hosts two zero-energy corner states, $|\psi_1\rangle$ and $|\psi_2\rangle$,
localized at the initial corner pair, as schematically illustrated in
Fig.~\ref{fig: BBH illustration}(b2).
For the superposition-state transfer, the localized corner-state basis is fixed by choosing each eigenvector to be real and requiring its largest-amplitude components in the corresponding corner region to be positive. 
Starting from the normalized superposition $(|\psi_1\rangle+|\psi_2\rangle)/\sqrt{2}$, the state adiabatically follows the instantaneous zero-mode subspace and evolves into a new superposition state of which the largest-amplitude components are localized at the other pair of corner regions. 
We define two fidelities, the 
gauge-invariant fidelity $\mathcal{F}_{\text{sub}}=|\langle\Psi(t)|\psi_3\rangle|^2+|\langle\Psi(t)|\psi_4\rangle|^2$, which characterizes whether the state is transferred into the desired target subspace, and 
$\mathcal{F}_{\text{total}}(t)=|\langle\Psi(t)|\Psi_{\text{tar}}^{\prime}\rangle|^2$, which verifies whether the intended coherent superposition is preserved during the state transfer. The target state is $|\Psi_{\text{tar}}^{\prime}\rangle=(|\psi_3\rangle+|\psi_4\rangle)/\sqrt{2}$, with [$|\psi_3\rangle$, $|\psi_4\rangle$] being the two eigenstates of the Hamiltonian $H_\text{tot}(t=T_2)=H+H_{A}+H_{B}$.
Exact Hamiltonian dynamics and noiseless circuit benchmarks for this superposition-transfer protocol are provided in the Supplemental Material~\cite{SuppleMat}, where both fidelities approach unity and the relative phase remains locked during the evolution.
We further demonstrate the transfer of a coherent corner-mode superposition [Fig.~\ref{fig:fig2_State_Transfer_Tianqi}(e)--\ref{fig:fig2_State_Transfer_Tianqi}(g)] on the same IBM quantum processor. The initial state at $t=0$, localized on two corners [$(1,1)$ and $(10,10)$] in Fig.~\ref{fig:fig2_State_Transfer_Tianqi}(e), is first prepared variationally~\cite{SuppleMat} and subsequently transferred to the opposite pair of corners [$(10,1)$ and $(1,10)$] at $t=T_2$ [Fig.~\ref{fig:fig2_State_Transfer_Tianqi}(g)]. 
The corresponding intermediate state is shown in Fig.~\ref{fig:fig2_State_Transfer_Tianqi}(f). 
As before, the circuits implementing each time-evolution step are compiled into shallow variational quantum circuits~\cite{SuppleMat}. Finally, using a unitary-rotation procedure similar to the single corner mode, we extract two transformed fidelities, $\mathcal{F}^*_{\rm total}$ and $\mathcal{F}^*_{\rm sub}$, which quantify, respectively, the preservation of the intended coherent superposition during state transfer and its arrival within the desired target subspace. As shown in Fig.~\ref{fig:fig2_State_Transfer_Tianqi}(h), both fidelities approach unity on a NISQ-era IBM quantum processor, confirming a coherent, switch-like reconfiguration of the corner-state occupation.  These results show that the subchiral-symmetry protocol enables not only deterministic single-mode transfer but also coherent manipulation within a degenerate corner-state manifold.


\textit{Conclusion and Discussion.}--
We have shown that subchiral symmetry can be used not only to classify or eliminate zero-energy corner modes, but also to control their dynamics. Our explicit results are based on the two-dimensional Benalcazar--Bernevig--Hughes model, where four symmetry sectors associated with the four zero-energy corner modes can be selectively broken by engineered intercell hoppings.  Building on this symmetry-resolved approach, we design adiabatic protocols for transferring either a single corner state or a coherent superposition of two corner states between designated lattice corners. 
In the latter case, the relative phase remains locked throughout the transfer, demonstrating genuine coherent transfer within a degenerate zero-mode manifold. 
We have shown that our control protocol can be executed on an IBM quantum processor with high fidelity, providing a hardware-level demonstration of subchiral-symmetry-enabled topological state manipulation. These results establish subchiral symmetry as an operational principle for programmable higher-order topological dynamics. They further suggest a general route for converting protected boundary modes into addressable quantum resources in other lattice geometries and quantum simulation platforms. Future directions involve extending this symmetry-resolved control principle beyond the current BBH model, including but not limited to other lattices that host higher-order topological physics~\cite{ezawa2018higher,xue2019acoustic,liu2023higher,koh2024realization,canyellas2024topological}, Floquet topological systems~\cite{huang2020floquet,hu2020dynamical,Zhu2021PRB,zhu2022time,huang2023topological,zhang2025observation,qian2025programmable}, and non-Hermitian platforms~\cite{kawabata2019symmetry,bergholtz2021exceptional,liu2019second,luo2019higher,li2020critical,okugawa2020second,wu2020wannier,liu2023anomalous,NHSE2025}, where subchiral or related subsystem symmetries may enable more exotic manipulation of corner modes.


\begin{acknowledgments}
{\bf Acknowledgments}: The authors thank P.~Song, Z.~Zou, and S.~Roy for helpful discussions. H.-T.~Ding thanks the Young Researcher Development Grant from Centre of Quantum Technologies, National University of Singapore (25-YRCDG-DHT).  T.~Chen acknowledges support by the National Research Foundation, Singapore, through the National Quantum Office, hosted in A*STAR, under the Advanced Quantum Algorithms and Solutions (AQAS) Funding Initiative (S25Q9DA001). 
J.~Gong also acknowledges support from the National Research Foundation, Singapore, through the National Quantum Office, hosted in A*STAR, under its Centre for Quantum Technologies Funding Initiative (S24Q2d0009).
We acknowledge the use of IBM Quantum services in this work. The views expressed are those of the authors and do not reflect the official policy or position of IBM or the IBM Quantum team.  
\end{acknowledgments}

\bibliography{references_new}
\end{document}